# Resonantly exited precession motion of three-dimensional vortex core in magnetic nanospheres

**Sang-Koog Kim[1,*], Myoung-Woo Yoo[1], Jehyun Lee[1,+], Ha-Youn Lee[1], Jae-Hyeok Lee[1], Yuri Gaididei[2], Volodymyr P. Kravchuk[2], and Denis D. Sheka[3]**

[1] National Creative Research Initiative Center for Spin Dynamics and Spin-Wave Devices, Nanospinics Laboratory, Research Institute of Advanced Materials, Department of Materials Science and Engineering, Seoul National University, Seoul 151-744, South Korea

[2] Bogolyubov Institute for Theoretical Physics, 03680 Kiev, Ukraine

[3] Taras Shevchenko National University of Kiev, 01601 Kiev, Ukraine

[*] sangkoog@snu.ac.kr

[+] Present address: Center of Semiconductor Research & Development, Gyeonggi-do 445-701, South Korea

## ABSTRACT

We found resonantly excited precession motions of a three-dimensional vortex core in soft magnetic nanospheres and controllable precession frequency with the sphere diameter $2R$, as studied by micromagnetic numerical and analytical calculations. The precession angular frequency for an applied static field $H_{DC}$ is given as $\omega_{MV} = \gamma_{eff} H_{DC}$, where $\gamma_{eff} = \gamma \langle m_\Gamma \rangle$ is the effective gyromagnetic ratio in collective vortex dynamics, with the gyromagnetic ratio $\gamma$ and the average magnetization component $\langle m_\Gamma \rangle$ of the ground-state vortex in the core direction. Fitting to the micromagnetic simulation data for $\langle m_\Gamma \rangle$ yields a simple explicit form of $\langle m_\Gamma \rangle \approx (73.6 \pm 3.4)(l_{ex}/2R)^{2.20 \pm 0.14}$, where $l_{ex}$ is the exchange length of a given material.

This dynamic behavior might serve as a foundation for potential bio-applications of size-specific resonant excitation of magnetic vortex-state nanoparticles, for example, magnetic particle resonance imaging.

# Introduction

The Larmor precession is a universal dynamic phenomenon in nature that represents the precession of a magnetic moment about a magnetic field at a characteristic Larmor frequency, which is expressed as $\omega_L = \gamma H$, where $\gamma$ is the gyromagnetic ratio and *H*, the static field strength. This type of precession plays very crucial roles in a rich variety of electron- or nuclei-spin-related dynamics such as electron-spin resonance, nuclear magnetic resonance, ferromagnetic resonance, and related magnetization dynamics.[1-6] Such dynamic fundamentals have been widely utilized in a significant number of applications, including material analysis,[1,4] bio-medical imaging,[7,8] and information recording in magnetic media.[9,10]

In this paper, we report the discovery of resonantly excited precession motions of a magnetic vortex core in soft magnetic nanoparticles of spherical shape,[11] but with totally different underlying physics from those for vortex motions so far reported.[12-17] We also were able to identify sphere-size-controllable precession angular frequency $\omega_{MV}$ and size-specific resonant excitations of nanoparticles bearing a magnetic vortex structure. We additionally determined, based on combined micromagnetic numerical and analytic calculations, that the size specificity of $\omega_{MV}$ originates from the variable effective gyromagnetic ratio with the sphere size that modifies the vortex structure inside spheres. Our results could provide a potential means of implementing size-specific resonant excitation of nanoparticles in bio-applications.[18]

# Results

**Ground states of nanospheres**

Figure 1a shows a nanosphere model of spherical symmetry. As described in Methods, we performed micromagnetic numerical calculations on Permalloy (Py, $Ni_{80}Fe_{20}$) nanoparticles of different diameters, $2R$ = 10 nm – 150 nm (see Methods). Figure 1b illustrates the ground states of the spheres obtained through relaxation from their saturated states in the +$x$ direction. For the $2R$ < 40 nm cases, uniformly magnetized single-domain states were obtained, whereas for the 50 nm $\leq 2R \leq$ 150 nm cases, single magnetic vortex states were well established. The vortex state of the $2R$ = 150 nm sphere, for example, was visualized by streamlines circulating around the vortex core oriented in the +$x$ direction. We noted that the region of the vortex core aligned in the +$x$ direction relative to the region of the in-plane circulating magnetizations varies markedly with $2R$, as indicated by the $x$-component of the local magnetization $m_x$ (= $M_x / M_s$) profiles in Fig. 1c. This dramatic variation is the result of a strong competition between the long-range dipolar and short-range exchange interactions in those nanospheres of such varying size.

**Resonantly excited precession motion of a vortex core in spheres**

Since the spherical symmetry of nanospheres does not lead to any magnetic shape anisotropy, when a sizable static field $H_{DC}$ is applied in the +$z$ direction, the vortex cores for 40 nm < $2R$ $\leq$ 150 nm start to reorient to the field direction, but with accompanying precession motions. This precession motion is different from the well-known gyration and even its higher-order modes of vortex cores in planar dots.[12-17] Although very weak spin waves are emitted inside the nanospheres, the vortex's spin configurations are maintained as a whole structure, because the field strength is sufficiently small. In the relaxation process, the core orientation converges in the field direction (+$z$-direction), reflecting the fact that the $m_x$ averaged over the entire volume of the sphere, <$m_x$>, undergoes decaying oscillation through its vortex-core precession (inset of Fig. 2a). The precession frequency was obtained by Fast Fourier Transformation (FFT) of the temporal <$m_x$> evolution for the different values of $2R$ and $H_{DC}$ (see Figs. 2a and 2b, respectively). In the cases of uniformly saturated particles ($2R$ = 10, 20, or 30 nm), the frequency was independent of $2R$, as determined by the Larmor frequency $f_L = (\gamma/2\pi) H_{DC}$.[19,20] By contrast, for the vortex-state spheres (40 nm ≤ $2R$ ≤ 120 nm), the precession frequency of a vortex core showed a strong variation with $2R$, as can be expressed by $f_{MV} = (\gamma_{eff}/2\pi) H_{DC}$, where $\gamma_{eff} (<\gamma)$ is the effective gyromagnetic ratio, which is variable with the sphere diameter.

In order to quantitatively elucidate the $\gamma_{eff}$-versus-$2R$ relation, we plotted the value of $f/H_{DC}$ as a function of $2R$, as compared with the average magnetization component over the sphere volume in the vortex-core orientation, $\langle m_\Gamma \rangle$, both of which were obtained from the micromagnetic simulations. As shown in Fig. 3, when $\gamma/2\pi$ = 2.8 (MHz/Oe) on the left axis is scaled to $\langle m_\Gamma \rangle$ = 1 on the right axis, both numerical values are in perfect quantitative agreement over the entire range of diameters studied, resulting in an explicit form of $\gamma_{eff}/\gamma = \langle m_\Gamma \rangle$ (for single-domain states, $\gamma_{eff} = \gamma$, because of $\langle m_\Gamma \rangle = 1$). Therefore, the precession frequency of a vortex core in nanospheres can be expressed as $f_{MV} = (\gamma/2\pi)\langle m_\Gamma \rangle H_{DC}$. This precession frequency cannot be explained by the gyration mode (or even by higher-order modes) of vortex cores in thin or thick film dots, and neither, consequently, by Thiele's equation.[12-17]

**Analytical derivation of vortex-core precession in nanospheres**

In order to gain physical insight into the $f_{MV} = (\gamma/2\pi)\langle m_\Gamma \rangle H_{DC}$ relation obtained from the micromagnetic simulations, we analytically derived vortex-core precession dynamics in nanospheres. In our modeling, a weak static field was applied in the +$z$ direction, which field sustained the rigid vortex structure in a certain potential, and thus allowed the initial ground-state vortex core to align in the +$z$ direction through the precession around the field direction along with certain damping. We used the local spherical reference frame on infinitesimal segments of the surface, where the unit vector of local magnetizations is expressed as $\mathbf{m} = (m_r, m_\theta, m_\varphi)$, $r$ is the radial distance, $\theta$ is the polar angle, and $\varphi$ is the azimuthal angle, as shown in Fig. 4a. Time-variable vortex-core orientation can be defined as a unit vector $\mathbf{\Gamma} \equiv (\sin\theta_0 \cos\varphi_0, \sin\theta_0 \sin\varphi_0, \cos\theta_0)$, as illustrated in Fig. 4b. Following the rigid vortex Ansatz, which agreed with the micromagnetic simulation results, local magnetizations inside a given sphere could be expressed as $m_r = f(r, \mathbf{\Gamma} \cdot \hat{\mathbf{r}})$ and $\Phi = -\arctan\{(\mathbf{\Gamma} \cdot \hat{\boldsymbol{\theta}})/(\mathbf{\Gamma} \cdot \hat{\boldsymbol{\varphi}})\}$, where $\Phi$ is the azimuthal angle of the magnetization in the local spherical reference frame (inset of Fig. 4a). Here we assume some general shapes of $m_r$ that are restricted by the condition $f(r,1) = -f(r,-1) = 1$ for all $r$ values. Since $m_r$ and $\Phi$ are canonically conjugated variables, the time evolution of the local magnetizations can be determined from the Landau-Lifshitz-Gilbert (LLG) equations[21,22] as

$$\dot{m}_r = -\frac{\gamma}{M_s}\frac{\delta E}{\delta \Phi} - \alpha(1-m_r^2)\dot{\Phi}, \tag{1a}$$

$$\dot{\Phi} = \frac{\gamma}{M_s}\frac{\delta E}{\delta m_r} + \frac{\alpha}{1-m_r^2}\dot{m}_r. \tag{1b}$$

By inserting the $m_r$ distribution function of the vortex's spin configuration into Eqs. (1a) and (1b), we finally obtained the governing equation for vortex-core precession motion,

$$\dot{\mathbf{\Gamma}} + \frac{\gamma}{M_s V}\mathbf{\Gamma} \times \frac{\partial E}{\partial \mathbf{\Gamma}} + \frac{\alpha}{V}\mathbf{\Gamma} \times \frac{\partial F}{\partial \dot{\mathbf{\Gamma}}} = 0, \qquad (2)$$

where $E$ is the total magnetic energy, $F$ is a dissipative functional ($F = \frac{V}{2}\int d\mathbf{r}[\sin^2\Theta(\frac{\partial \Phi}{\partial \mathbf{\Gamma}}\cdot\dot{\mathbf{\Gamma}})^2 + (\frac{\partial \Theta}{\partial \mathbf{\Gamma}}\cdot\dot{\mathbf{\Gamma}})^2]$), and $V = \frac{4}{3}\pi R^3$ is the sphere volume. The first, second and third terms in Eq. (2) correspond to the gyrotropic, potential energy and damping terms, respectively. The total energy $E$ under a weak magnetic field applied along the $z$-axis, $\mathbf{H} = H_{DC}\hat{\mathbf{z}}$, can be expressed simply as $E_H = -\hat{\mathbf{z}}\cdot\mathbf{\Gamma}(t)H_{DC}VM_s\langle m_\Gamma\rangle$, where $\langle m_\Gamma\rangle$ is rewritten as $\langle m_\Gamma\rangle = \frac{1}{M_s V}\int d\mathbf{r}(\mathbf{M}\cdot\mathbf{\Gamma})$. Eq. (2) expresses the precession motion of vortex cores in collective spin dynamics; it differs from Thiele's equation to describe the gyration of vortex cores in planar dot systems.

By inserting $E_H$ into Eq. (2) and assuming negligible damping, the precession frequency of a rigid vortex core can be given as $\partial\varphi_0/\partial t = 2\pi f_{MV}$ with $f_{MV} = (\gamma/2\pi)\langle m_\Gamma\rangle H_{DC}$. Consequently, we obtained the effective gyromagnetic ratio of the motion of a vortex in a given nanosphere as $\gamma_{eff} = \gamma\langle m_\Gamma\rangle$. This analytic form provides a clear physical insight into $2R$-dependent $f_{MV}$, because $\langle m_\Gamma\rangle$, as indicated in the micromagnetic simulation results, varies with $2R$. Here we note that the eigenfrequency of a single vortex in cylindrical dots is known to vary with the aspect ratio of thickness $L$ to $R$.[12,15-17] However, the underlying physics of the size-dependent change in the precession frequency of the vortex core in nanospheres is totally different from that of the vortex gyration in planar disks, though both apparently show core-oscillation phenomena.

**Dependence of $\langle m_\Gamma \rangle$ on sphere's diameter and constituent material parameters**

Next, it is necessary to quantify how $\langle m_\Gamma \rangle$ varies with $2R$. We estimated, from further micromagnetic numerical calculations, the quantitative relation between $\langle m_\Gamma \rangle$ and $2R$ within the $2R = 50 - 200$ nm range for the different material parameters of both $M_s$ and $A_{ex}$. Figure 5 reveals that $\langle m_\Gamma \rangle$ is given as $\langle m_\Gamma \rangle \approx \eta [A_{ex}/(2\pi M_s^2)]^{1.09 \pm 0.05} (2R)^{-2.21 \pm 0.13}$ with $\eta = 73.6 \pm 3.4$. According to the relation $l_{ex} = \sqrt{A_{ex}/(2\pi M_s^2)}$,[1] $\langle m_\Gamma \rangle$ can be simplified as $\langle m_\Gamma \rangle \approx (73.6 \pm 3.4)(l_{ex}/2R)^{2.20 \pm 0.14}$. This explicit form provides a simple and reliable estimation of $\langle m_\Gamma \rangle$ for a given value of $2R$ and a given material of $l_{ex}$, though there is yet no concrete model matching the form. We also note that, based on the single-domain states of $\langle m_\Gamma \rangle = 1$, the critical size for transition from a single domain to a vortex state[1,23] can be simply estimated as $2R_c = 7.06\, l_{ex}$. For example, the critical diameter, $2R_c = 37.3$ nm for Py, was in good agreement with that obtained from the simulation results shown in Fig. 1. As quantitatively interpreted, the strong variation of $\langle m_\Gamma \rangle$ versus $2R$ for a given material is related to the competition between the short-range, strong exchange interaction and long-range, but relatively weak dipolar interaction in nanospheres of given dimensions.

**Size-specific resonant excitations**

As an application of the aforementioned fundamental dynamics, we could activate magnetic nanoparticles of a specific size by tuning the frequency of an applied AC field to the $f_{MV}$ of a sphere of a given diameter and material. In this modeling, an external AC field and a static field were given by $\mathbf{H}_{AC} = H_{AC}\sin(2\pi f_{AC}t)\hat{\mathbf{y}}$ and $\mathbf{H}_{DC} = H_{DC}\hat{\mathbf{z}}$, respectively, with sufficiently small values of $H_{AC}$ = 10 Oe and $H_{DC}$ = 100 Oe to avoid deformation of the initial vortex structures in the Py spheres. Figure 6a shows the oscillation of the core orientation $\theta_0$ from the +z direction during the precession process for 2$R$ = 60 nm ($f_{MV}$ = 95 MHz), as excited by $f_{AC}$ = 91, 95 and 99 MHz. The oscillation of $\theta_0$ was hardly observable for the cases where $f_{AC}$ was far from $f_{MV}$, whereas it was very large for the case of $f_{AC} = f_{MV}$, that is, at resonance. The resonantly excited precession leads even to vortex-core reversals between $\theta_0 = +\pi$ and 0, as such reversals in planar disks occur periodically by linearly oscillating fields or currents applied on the disks' plane under the resonance condition.[3,24] The oscillation of $\theta_0$ represents a transfer of the external magnetic field to a magnetic sphere via the absorption of the Zeeman energy and subsequent emission to another form. The maximum energy absorption can be defined by the first maximum energy increment, $\Delta E_1$, as noted in Fig. 6a. Figure 6b plots $\Delta E_1$ versus $f_{AC}$ for different sphere diameters.[25] For each diameter, the maximum peak height $\Delta E_{max}$ in the $\Delta E_1$-versus-$f_{AC}$ curves was obtained under the corresponding resonance condition. All of the curves were well separated from each other, indicating reliable size-specific excitation of the magnetic particles. For example, the difference in $f_{MV}$ between the 50 and 60 nm particles was about 50 MHz, which is sufficiently large compared with the full width at half maximums of both particles, 6.6 and 9.9 MHz, respectively.

In Fig. 6c are shown the $\Delta E_{max}$-versus-2$R$ curves for comparison between the simulation data (solid circles) and the analytical form (lines) of the Zeeman energy, $\Delta E_{max} = 2H_{DC}VM_s \langle m_\Gamma \rangle$, where $\langle m_\Gamma \rangle = 1$ for single-domain states or $\langle m_\Gamma \rangle \approx (73.6 \pm 3.4)(l_{ex}/2R)^{2.20 \pm 0.14}$ for vortex states. The simulation and analytical calculation agreed very well, as can be seen. The analytical calculation clearly shows that the magnetic energy absorption varies with $(2R)^3$ and $(2R)^{0.8}$ for the single-domain and vortex states, respectively. These results suggest that the magnetic energy absorption can be maximized by tuning $f_{AC}$ to the resonance frequency of a given-diameter particle. This effect is made possible through size-specific resonance, size-selective activation and corresponding detection of the magnetic nanoparticles of a vortex state.

**Discussion**

We discovered, by micromagnetic numerical calculations, not only the resonantly excited precession motion of a vortex core in nanospheres and its size-dependent precession frequency, but also its physical origin, based on the size effect on the effective gyromagnetic ratio in collective spin dynamics analytically derived. This finding paves the way for size-selective activation and/or possible detection of magnetic nanoparticles by application of extremely low-strength AC fields tuned to the resonant frequency of a given diameter and material. These results, notably, would be applicable to magnetic particle resonance imaging (MPRI) and bio-applications.

**Methods**

In our micromagnetic numerical calculations, the FEMME code (version 5.0.8)[26] was used to numerically calculate the motions of the magnetizations of individual nodes (mesh size: ≤ 4 nm) interacting with each other via exchange and dipolar interactions at the zero temperature, as based on the LLG equation.[21,22] The surfaces of the model spheres were discretized into triangles of roughly equal area using Hierarchical Triangular Mesh (HTM), as shown in Fig. 1a, in order to prevent irregularity-incurred numerical errors.[27] The chosen material parameters corresponding to Py were as follows: saturation magnetization $M_s$ = 860 emu/cm$^3$, exchange stiffness $A_{ex}$ = 1.3 × 10$^{-6}$ erg/cm, damping constant $\alpha$ = 0.01, $\gamma / 2\pi$ = 2.8 MHz/Oe, and zero magnetocrystalline anisotropy for the soft ferromagnetic Py material.

**References**

1. O'Handley, R. C. *Modern Magnetic Materials: Principles and Applications*. (Wiley, 1999).

2. Hillebrands, B. & Thiaville, A. *Spin Dynamics in Confined Magnetic Structures III*. (Springer, 2006).

3. Lee, K. S., Guslienko, K. Y., Lee, J. Y. & Kim, S. K. Ultrafast vortex-core reversal dynamics in ferromagnetic nanodots. *Phys. Rev. B* **76**, 174410 (2007).

4. Coey, J. M. D. *Magnetism and Magnetic Materials*. (Cambridge University Press, 2010).

5. Kammerer, M. *et al.* Magnetic vortex core reversal by excitation of spin waves. *Nat. Commun.* 2:279 doi: 10.1038/ncomms1277 (2011).

6. Yoo, M. W., Lee, J. & Kim, S. K. Radial-spin-wave-mode-assisted vortex-core magnetization reversals. *Appl. Phys. Lett.* **100**, 172413 (2012).

7. Trabesinger, A. Imaging techniques - Particular magnetic insights. *Nature* **435**, 1173-1174 (2005).

8. Webb, A. G. *Introduction to Biomedical Imaging*. (Wiley-IEEE Press, 2002).

9. Devolder, T. & Chappert, C. Precessional switching of thin nanomagnets: analytical study. *Eur. Phys. J. B* **36**, 57-64 (2003).

10. Kim, S. K., Lee, K. S., Yu, Y. S. & Choi, Y. S. Reliable low-power control of ultrafast vortex-core switching with the selectivity in an array of vortex states by in-plane circular-rotational magnetic fields and spin-polarized currents. *Appl. Phys. Lett.* **92**, 022509 (2008).

11. Kim, M. K. *et al.* Self-assembled magnetic nanospheres with three-dimensional magnetic vortex. *Appl. Phys. Lett.* **105**, 232402 (2014).

12. Guslienko, K. Y. *et al.* Eigenfrequencies of vortex state excitations in magnetic submicron-size disks. *J. Appl. Phys.* **91**, 8037-8039 (2002).

13. Boust, F. & Vukadinovic, N. Micromagnetic simulations of vortex-state excitations in soft magnetic nanostructures. *Phys. Rev. B* **70**, 172408 (2004).

14. Yan, M., Hertel, R. & Schneider, C. M. Calculations of three-dimensional magnetic normal modes in mesoscopic permalloy prisms with vortex structure. *Phys. Rev. B* **76**, 094407 (2007).

15. Zarzuela, R., Chudnovsky, E. M. & Tejada, J. Excitation modes of vortices in submicron magnetic disks. *Phys. Rev. B* **87**, 014413 (2013).

16. Zarzuela, R., Chudnovsky, E. M., Hernandez, J. M. & Tejada, J. Quantum dynamics of vortices in mesoscopic magnetic disks. *Phys. Rev. B* **87**, 144420 (2013).

17. Ding, J., Kakazei, G. N., Liu, X. M., Guslienko, K. Y. & Adeyeye, A. O. Higher order vortex gyrotropic modes in circular ferromagnetic nanodots. *Sci Rep-Uk* **4**, 4796 (2014).

18. Mornet, S., Vasseur, S., Grasset, F. & Duguet, E. Magnetic nanoparticle design for medical diagnosis and therapy. *J. Mater. Chem.* **14**, 2161-2175 (2004).

19. Kittel, C. On the Theory of Ferromagnetic Resonance Absorption. *Phys. Rev.* **73**, 155-161 (1948).

20. For the saturation magnetization, the resonance frequency of an ellipsoid with the demagnetizing factors $N_x$, $N_y$ and $N_z$ under a static magnetic field applied in the z-direction is given as $\omega = \gamma\sqrt{\left[H_z+(N_y-N_z)M_z\right]\cdot\left[H_z+(N_x-N_z)M_z\right]}$. Therefore, it can be concluded that the demagnetization factors (i.e., particle geometries) can influence the resonant frequency. In our nanosphere magnets case, the Kittel equation was simply rewritten $\omega = \gamma H_z$, because the demagnetization factors of the sphere were given as $N_x = N_y = N_z = 4\pi/3$.

21. Landau, L. D. & Lifshitz, E. M. On the theory of the dispersion of magnetic permeability

in ferromagnetic bodies. *Phys. Zeitsch. der Sow.* **8**, 153-169 (1935).

22. Gilbert, T. L. A phenomenological theory of damping in ferromagnetic materials. *IEEE T. Magn.* **40**, 3443-3449 (2004).

23. Bertotti, G. *Hysteresis in Magnetism: For Physicists, Materials Scientists, and Engineers*. (Academic Press, 1998).

24. Yamada, K. *et al.* Electrical switching of the vortex core in a magnetic disk. *Nat. Mater.* **6**, 269-273 (2007).

25. Chen, S. W., Chiang, C. L. & Chen, C. L. The influence of nanoparticle size and external AC magnetic field on heating ability. *Mater. Lett.* **67**, 349-351 (2012).

26. Suess, D., & Schrefl, T. FEMME: Finite Element MicroMagnEtics 5.0.8 (SuessCo, http://suessco.com/).

27. Szalay, A. S., Gray, J., Fekete, G., Kunszt, P. Z., Kukol, P., & Thakar, A. Indexing the Sphere with the Hierarchical Triangular Mesh. e-print arXiv:cs/0701164 (2007).

## ACKNOWLEDGMENTS

This research was supported by the Basic Science Research Program through the National Research Foundation of Korea funded by the Ministry of Science, ICT & Future Planning (grant no. 2014001928).

## Author contributions statement

S.-K.K. and J.L. conceived the main idea and planned the micromagnetic simulation. M.-W.Y., J.L., H.-Y. L. and J.-H.L. performed the micromagnetic simulations and M.-W.Y., J.L., analyzed the data. Y.G., V.P.K and D.D.S. derived analytical expressions for the precession motion of vortex core. S.-K.K wrote the manuscript and all the coauthors commented on the manuscript.

## Additional information

**Competing financial interests**: The authors declare no competing financial interests

**Figure Legends**

**Figure 1. Ground-state magnetization configurations of Py nanospheres according to the diameter.** (a) Finite-element sphere model for diameter $2R$ = 30 nm. (b) Ground-state magnetization configurations of Py nanospheres for different $2R$ values as indicated: upper, viewed from positive $z$-direction and sliced across $x$-$y$ plane; lower, viewed from positive $x$-direction and sliced across $y$-$z$ plane. The color represents the $x$-component of the local magnetizations, $m_x = M_x / M_s$ (see the color bar). The arrows inside the sphere of $2R$ = 150 nm represent the local curling magnetizations. (c) $m_x$ profiles along $y$ axis for different diameters. The inset shows the $m_x$ profiles versus the normalized distance for each diameter.

**Figure 2. Precession frequency of Py nanospheres as functions of (a) $2R$ and (b) $H_{DC}$ applied in +$z$ direction (perpendicularly to initial vortex-core orientation).** The inset in (a) shows the $<m_x>$ oscillation versus time, for a sphere of $2R$ = 80 nm. In (a), uniform single-domain (SD) and vortex states are distinguished at about $2R$ = 37 nm by the gray color. The symbols indicate the micromagnetic numerical calculations, with corresponding lines drawn by eye. In (b), the lines are the results of linear fits for the individual diameters, as indicated.

**Figure 3. Precession frequency normalized by $H_{DC}$ (circles) and $\langle m_\Gamma \rangle$ (crosses) obtained from micromagnetic numerical calculations.** The value of $\gamma/2\pi$ = 2.8 (MHz/Oe) on the left axis is scaled to $\langle m_\Gamma \rangle$ = 1 on the right axis. The different colors of the circle symbols indicate the numerical data for different sphere diameters, as indicated by the colors shown in

Fig. 2a. The solid curve is the result of a numerical calculation of the analytical form of $\langle m_\Gamma \rangle \approx (73.6 \pm 3.4)(l_{ex}/2R)^{2.20\pm 0.14}$.

**Figure 4. Model for analytical derivations.** (a) Definition of spherical coordinates and local spherical reference frame (colored surface) for local magnetization **m**. (b) Schematic of model sphere wherein single rigid vortex core is pointed in direction of $\theta_0$ and $\varphi_0$, as defined by the polar and azimuthal angle coordinates.

**Figure 5. Calculation of** $\langle m_\Gamma \rangle$ **for indicated values of** $2R$**,** $A_{ex}/A_{ex,Py}$ **and** $M_s/M_{s,Py}$**.** For each graph, one parameter is fixed: (a) $A_{ex}/A_{ex,Py} = 1$, (b) $M_s/M_{s,Py} = 1$, (c) and (d) $2R = 100$ nm. In both (a) and (b), $2R$ is in the 50 – 200 nm range. All of the symbols were obtained from the micromagnetic simulation results. The lines indicate linear fits.

**Figure 6. Total magnetic energy variation.** (a) Total magnetic energy and polar angle of core orientation $\theta_0$ during excitations of vortex core in sphere of $2R = 60$ nm by oscillating fields of $H_{AC} = 10$ Oe with different field frequencies ($f_{AC} = 91$, 95 and 99 MHz) under static field of $H_{DC} = 100$ Oe applied on +z-axis. (b) Plot of $\Delta E_1$ versus $f_{AC}$ in $f_{AC} = 25 – 310$ MHz range. (c) Maximum absorption energy $\Delta E_{max}$ versus $2R$, calculated from micromagnetic simulations (solid circles) and analytical form (lines) of $\Delta E_{max}$ described in text.

**Figure 1**

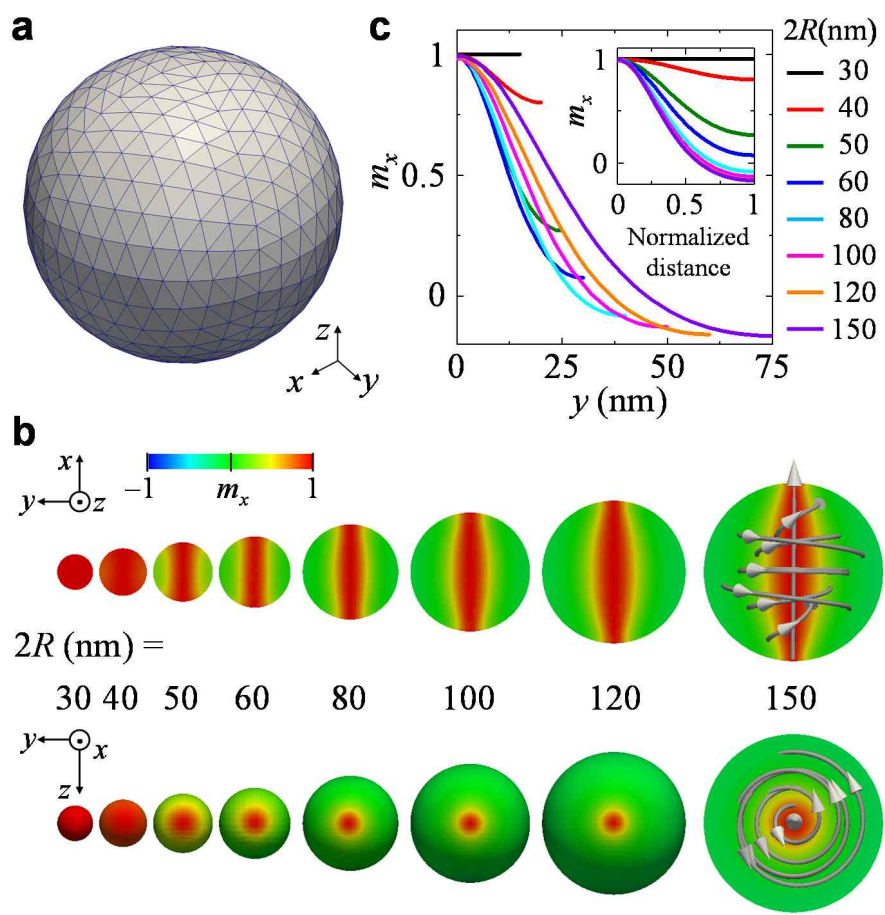

**Figure 2**

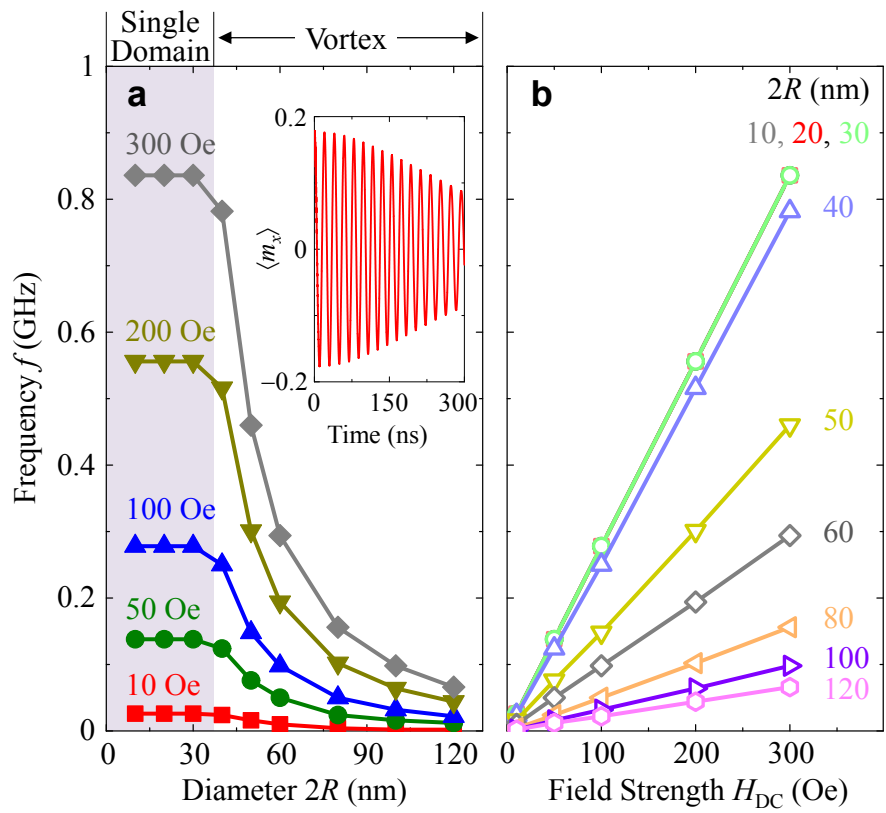

**Figure 3**

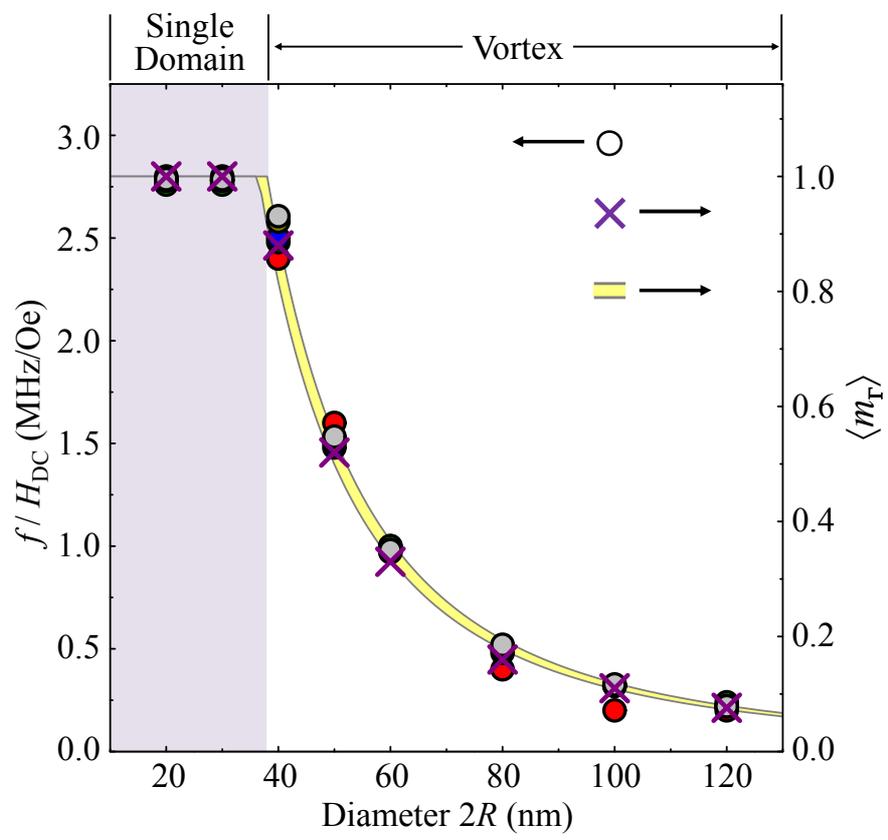

**Figure 4**

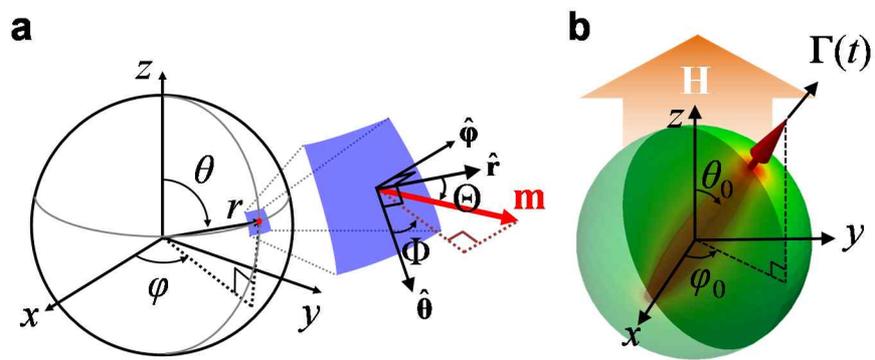

**Figure 5**

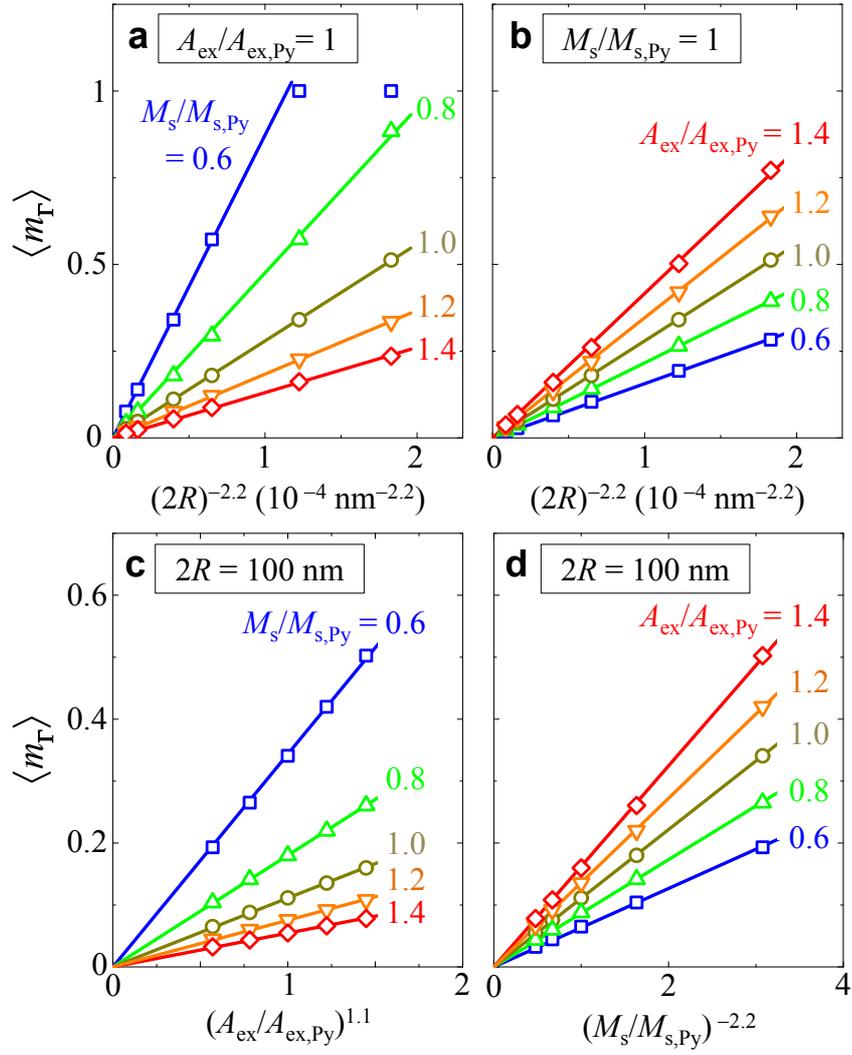

**Figure 6**

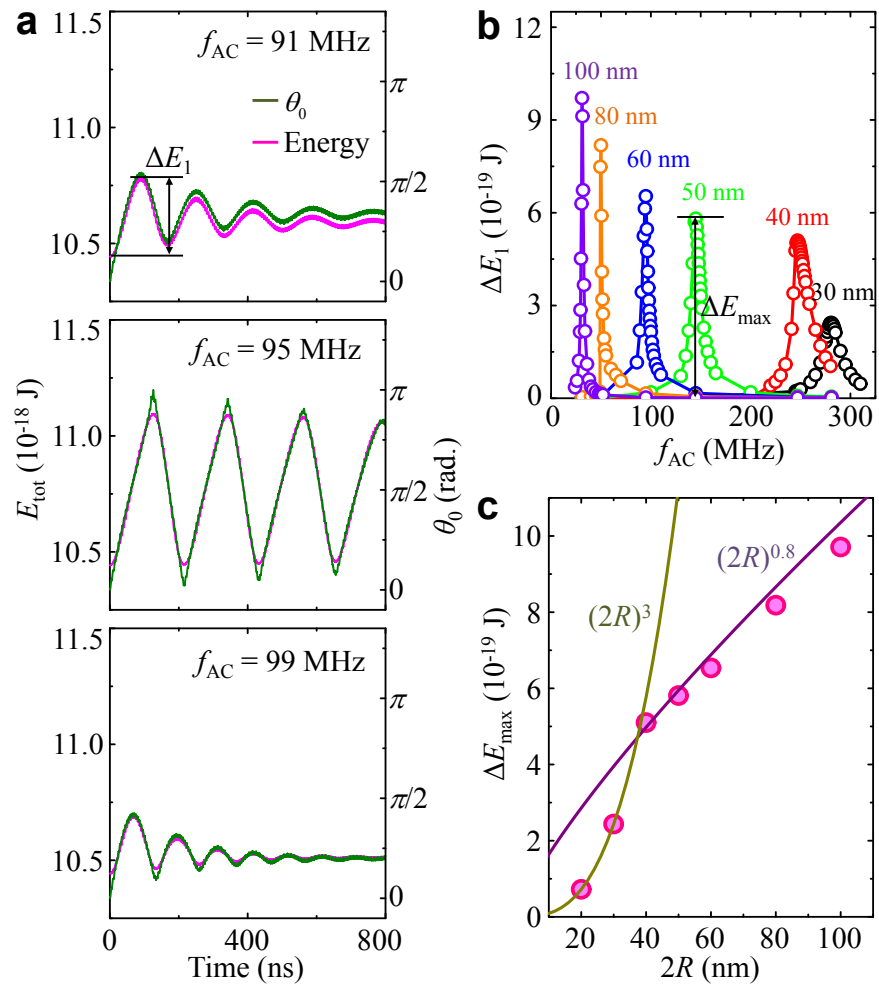